\documentclass[referee]{raa}            
\usepackage{graphicx}             
\usepackage[caption=false,font=footnotesize]{subfig}
\begin{document}
   \title{Antenna system characteristic and solar radio burst observation
}

   \volnopage{Vol.0 (200x) No.0, 000--000}      
   \setcounter{page}{1}          

   \author{Sha Li
      \inst{1,2,3}
   \and Yi-hua Yan
      \inst{1,2}
   \and Zhi-jun Chen
      \inst{1,2}
      \and Wei Wang
      \inst{1,2}
      \and Dong-hao Liu
      \inst{1}
   }

   \institute{Key Laboratory of Solar Activity, National Astronomical Observatories, Chinese Academy of Sciences;  Datun Road 20A, Chaoyang District, Beijing 100012, China {\it lisha1400@bao.ac.cn}\\
        \and
             Key Laboratory of Radio Astronomy, Chinese Academy of Science\\
         \and
             University of Chinese Academy of Sciences 10049\\
   }

   \date{Received~~2015 month day; accepted~~2015~~month day}

\abstract{ Chinese Spectral Radio Heliograph (CSRH) is an advanced aperture synthesis
solar radio heliograph, developed by National Astronomical
Observatories, Chinese Academy of Sciences independently. It
consists of 100 reflector antennas, which are grouped into two
antenna arrays (CSRH-I and CSRH-II) for low and high frequency bands
respectively. The frequency band of CSRH-I is 0.4-2GHz and for CSRH-II, the frequency band is 2-15GHz. In the antenna and feed system, CSRH uses an Eleven feed to
receive signals coming from the Sun, the radiation pattern with lower side lobe and back lobe of the feed
is well radiated. The characteristics of gain $ G $  and antenna noise
temperature $ T $ effect the quality of solar radio
imaging. For CSRH, measured $ G $ is larger than 60 dBi and $ T $ is less than 120K, after CSRH-I was established, we have successfully captured a solar radio burst between 1.2-1.6GHz on November 12, 2010 through this instrument and this event
was confirmed through the observation of Solar Broadband Radio Spectrometer (SBRS) at 2.84GHz and Geostationary Operational Environmental Satellite (GOES). In
addition, an image obtained from CSRH-I clearly reveals the profile of the
solar radio burst. The other observational work is the imaging of Fengyun-2E geosynchronous satellite
which is assumed to be a point source. This data processing method indicates that, the method of deleting errors about dirty image could be used for processing other surface sources.
\keywords{Reflector antenna :Noise Temperature : Eleven
antenna : Low cross polarization : Aperture synthesis
}
}

   \authorrunning{S. Li, Y.-H. Yan, Z.-J. Chen, W. Wang \&D.-H. Liu}            
   \titlerunning{Antenna system characteristic and solar radio burst observation}  

   \maketitle

%
%
\section{Introduction}           
\label{sect:intro}

In solar radio observations, the Sun exhibits a
variety of large dynamic phenomena in different frequencies. At the
same time, it transmits constant energy to the Earth (Takano ~\cite{Takano}).
These phenomena reveal the links between solar astronomy and other
branches of physics. For example, magnetohydrodynamics(MHD)and plasma physics can be well explained
through solar observations. The fundamental concept behind MHD shows that the magnetic fields can induce currents in a moving conductive fluid, which creates forces on the fluid and also changes the magnetic field itself. The plasma is composed of ions and electrons, it is the key to understand the propagation of radio wave coming from the Sun. CSRH observations with high spatial resolution could provide important diagnosing tools on the magnetic field, solar radio density, plasma temperature,etc (Yan~\cite{Yan}). Imaging spectroscopy over centimeter and decimeter wavelength range are important for CSRH to explain fundamental problems such as energy release, particle acceleration and particle transport.

In view of the available solar radio astronomical instruments all
over the world, two main kinds of instruments are used in solar
observations. One is radio heliograph at single or discrete
frequencies, for example, Nancay Radio Heliograph (NRH) observes the Sun at 150,
164, 237, 327 and 410MHz, Nobeyama Radio Heliograph (NoRH)
(Nakajima~\cite{Nakajima}) observes the Sun at 17 and 34GHz and Siberian Solar Radio
Telescope(SSRT)(Uralov~\cite{Uralov}) observes the Sun at 5.7GHz (Les~\cite{Les}). The other instruments are spectrometers
at certain frequency band and frequency points, such as Solar Broadband Radio
Spectrometer (SBRS) at 1-2GHz, 2.6-3.8GHz, 5.2-7.6GHz, 2.84GHz. Although scientists
have developed some good theoretic models from the observations of
these instruments, there were still a lot of phenomena that could
not be explained by the current observations and theories. So a radio heliograph which could give
solar images in ultra wide band width is required. Now, an undergoing CSRH
with high temporal, high spatial and high
spectral resolution will create radio images in ultra wide band width. It could
be expected to provide better observations explaining these
phenomena(Chen~\cite{Chen}).

 The aim of CSRH is to provide a new tool observing solar radio emissions, including radio bursts from primary energy release sites of the solar energetic events such as flares and coronal mass ejections (CMEs). Solar radio bursts are rich with different types during solar flares, which are believed due to the sudden energy release process of the solar magnetic field topological re-organization or through the magnetic reconnection. CSRH will make full disk of solar radio images with multiple frequency channels with 25ms cadence in the measurement.

Errors in radio heliograph, such as correlation-based error,
antenna pointing precision error, the receiver output
noise error. Because of so many errors, we have to calibrate the phase and amplitude errors of each output signal. The amplitude calibration is performed by comparing and equalizing the signal
levels in the element of antenna array. With consideration of the instrument's errors, the measurement of phase
calibration between different channels is to make use of celestial radio source from this theory, and the phase corrections are applied to the actual solar observations. As we know,
the sensitivity of a point source is proportional to the effective
collecting area of the telescope, this area is decided by one antenna multiplied by the number of the antennas, in fact, the Sun is larger than the beam of the individual antennas (Liu~\cite{Gary}), so a number of pointing directions are used.

To evaluate CSRH antenna system, the gain and noise temperature of
the antenna, the sensitivity of the receiver should be measured.
This paper presents a method for computing the characteristics of
CSRH, and provides the results of the antenna gain $ G $ and antenna temperature $ T $. This paper also
gives two observation work, one is solar radio burst observation
and the other is satellite source imaging.
The solar radio burst was observed by
CSRH-I 5-element array on December 12, 2010. This instance was also
captured by SBRS at 2.84GHz and GOES in X-ray. Another result came from observing the Fengyun-2E
geosynchronous satellite at the height of nearly 35600km. The output
signal from each antenna was calibrated with
reference to the known standard satellite source. Then, we produced
the images by using Common Astronomy Software Application (CASA)
which is an image layout software used for aperture synthesis
telescope.

The content of this paper is arranged as follows.
Section II gives the description of CSRH along with the
measured and simulated radiation patterns. In section III,
the antenna noise temperature and gain of the system are calculated in detail. Section IV gives images of the solar radio burst and satellite point source, which are drew by using the calibrated data of CSRH. Finally, the
conclusions are given in section V.


\section{Antenna system}
\subsection{About CSRH}
\begin{figure}
 \centering
  \includegraphics[width=0.8\textwidth]{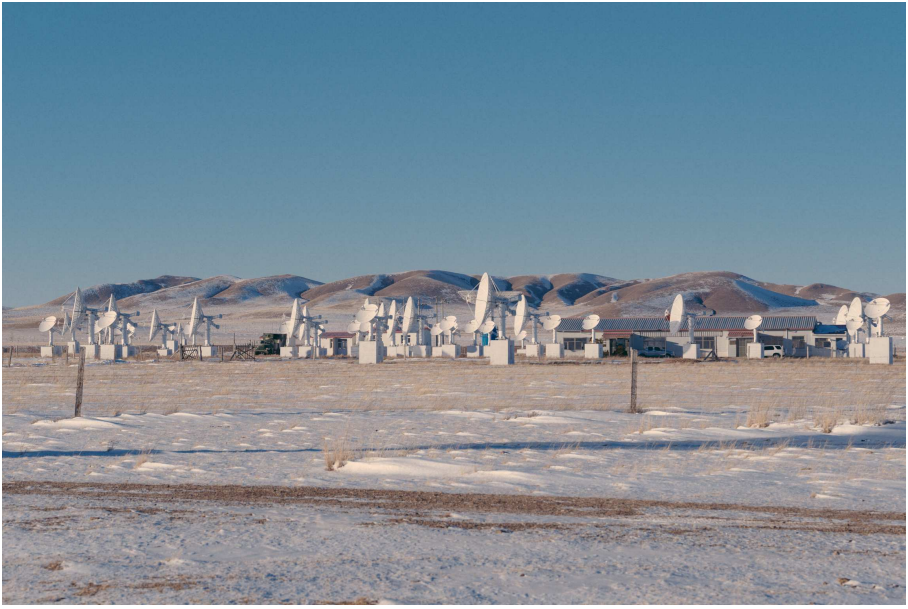}\\
  \caption{A view of CSRH}
  \label{figure:1}
\end{figure}

CSRH is a radio interferometer
which contains 100 parabolic reflector antennas. CSRH-I consists of
40 4.5-m-diameter antennas operating between 0.4-2GHz, CSRH-II
consists of 60 2-m-diameter antennas from 2
to 15GHz. (Figure~\ref{figure:1}) shows the photo of constructed
CSRH, which includes 3.87 hectares and locates in Inner Mongolia, 400km away from Beijing. The exact geographical coordinates of the central antenna is located at 115 degrees 15 minutes 1.8 seconds east longitude and 42 degrees 12 minutes 42.6 seconds north latitude. The antenna configuration (Yan~\cite{Yan}) is a non-redundant array with good (${u}$,${v}$) coverage. The baseline vector has components (${u}$,${v}$,${w}$), where ${w}$ points in the direction of interest. ${u}$,${v}$,${w}$ are measured in wavelengths at center frequency of the RF signal band, and in the directions towards the East, the North, and the phase tracking center respectively. For (${u}$,${v}$) coverage, if we assume the spatial frequencies as follows:
\begin{equation}
  \overrightarrow {\rm{f}} {\rm{ = }}{\rm{(u,v)}}
\end{equation}
which are the conjugated coordinates of the spatial coordinates (x,y) in the image plane. While (x,y) measure angles, usually expressed in arcseconds, spatial frequencies measure distance in the incident wavefront measured in wavelength units, they are usually expressed in $ arcsec^{-1} $, shown in (Equation~\ref{eq1}),
\begin{equation}
   \overrightarrow {\rm{f}} {\rm{ = }}{\rm{(u,v)}}{\rm{ = }}\frac{1}{\lambda }(\Delta {\rm{X,}}\Delta {\rm{Y}})
   \label{eq1}
\end{equation}

\begin{figure*}
\centering
  \includegraphics[width=0.65\textwidth]{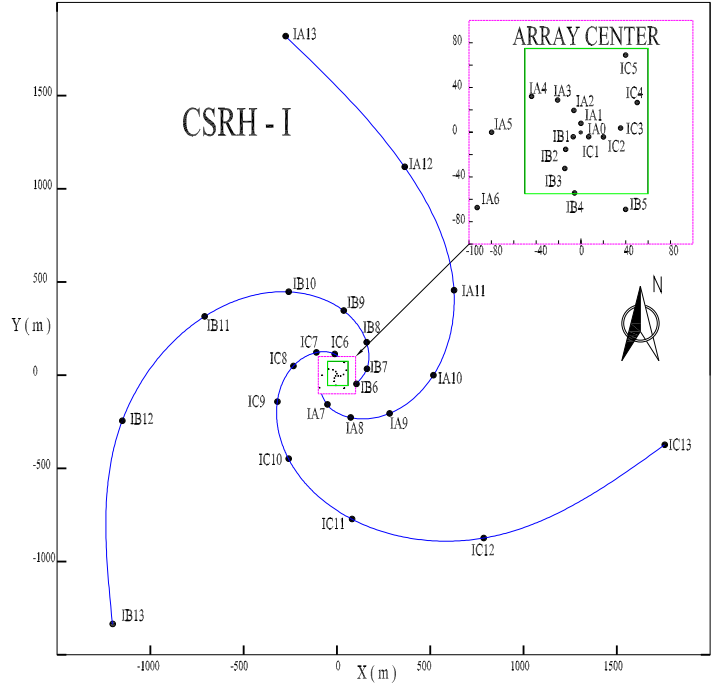}
\caption{The 2D configuration of CSRH-I, the antennas are arranged in three spiral arms (named A axis, B axis and C axis respectively), every axis has 13 antennas, IA0 is in the center of the antenna array, its correspondent coordinate is (0,0)m, the locations of other antennas are reference to IA0 antenna, the black circle shown in this figure is the location of each antenna}
\label{figu:1}       
\end{figure*}

\begin{figure*}
\centering
  \includegraphics[width=0.8\textwidth]{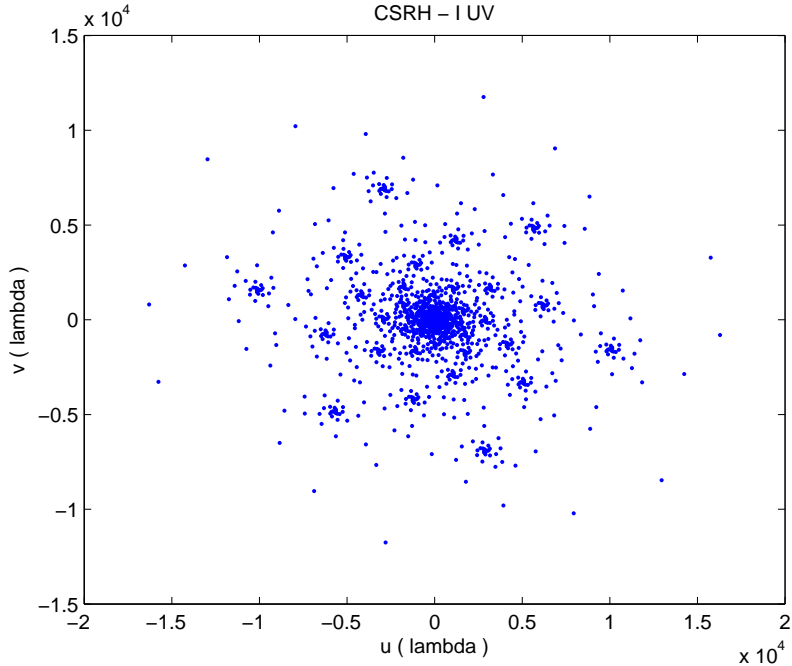}
\caption{(${u}$,${v}$) coverage of CSRH-I. The blue points correspond to the locations of each (${u}$,${v}$) coverage of the source image. The number of the baselines is 780 for this radio heliograph. A stellar interferometer measures only one visibility per baseline.}
\centering
\label{fig:1}       
\end{figure*}

During the imaging process, the incoming wavefront is spatially sampled by the radio heliograph, we need to make our measurements in a plane and measured in terms of the wavelength. When we observe the Sun, the sampling of the incident wavefront is no longer continuous and depends on the array configuration. The Fourier components of the object are measured at different spatial frequencies. (Figure~\ref{figu:1}) shows the arrangement of CSRH-I. (Figure~\ref{fig:1}) gives the (${u}$,${v}$) coverage (Lindsay~\cite{Lindsay}) of (Figure~\ref{figu:1}). All the antennas are arranged in three spiral arms (named A axis, B axis and C axis) respectively.

For multiple-element antenna arrays, it is convenient to specify the antenna positions relative to the reference point measured in a Cartesian coordinate system, a system with axes pointing towards hour-angle h and declination $ \delta $
 equal to (h=0, $ \delta=0 $) for X, (h=-$ {6^h} $, $ \delta =0 $) for Y, and ($ \delta  = 90 \deg $) for Z. If we
 assume $ \L_X $, $ \L_Y $ and $ \L_Z $ are the corresponding coordinate differences for two antennas, the baseline
 components (${u}$,${v}$,${w}$) are given by (Equation~\ref{UV}):
\begin{equation}\label{UV}
\left( \begin{array}{l}
u\\
v\\
w
\end{array} \right) = \frac{1}{\lambda }\left( \begin{array}{l}
\begin{array}{*{20}{c}}
{}&{\sin {H_0}}&{\begin{array}{*{20}{c}}
{}&{}
\end{array}\begin{array}{*{20}{c}}
{}&{\sin {H_0}}&{\begin{array}{*{20}{c}}
{}&{}
\end{array}\begin{array}{*{20}{c}}
{}&{}
\end{array}}&0
\end{array}}
\end{array}\\
 - \sin {\delta _0}\cos {H_0}\begin{array}{*{20}{c}}
{}&{\sin {\delta _0}\sin {H_0}\begin{array}{*{20}{c}}
{}&{\cos {\delta _0}}
\end{array}}
\end{array}\\
\cos {\delta _0}\cos {H_0}\begin{array}{*{20}{c}}
{}&{ - \cos {\delta _0}\sin {H_0}\begin{array}{*{20}{c}}
{}&{\sin {\delta _0}}
\end{array}}
\end{array}
\end{array} \right)\left( \begin{array}{l}
{L_X}\\
{L_Y}\\
{L_Z}
\end{array} \right)
\end{equation}
where $ {H_0} $ and $ \delta _0 $ are the hour-angle and declination of the phase reference position, and $ \lambda $ is the wavelength corresponding to the center frequency of the receiving system. The elements in the transformation matrix in (Equation~\ref{UV}) are the direction cosines of the (${u}$,${v}$,${w}$) axis relative to (X,Y,Z) axes. Thus as the interferometer observes a point source on the celestial sphere, the rotation of the Earth causes the ${u}$ and ${v}$ components of the baseline to trace out an elliptical locus. This ellipse is simply the projection onto the (${u}$,${v}$) plane of the circular locus traced out by the tip of the baseline vector, and at any instant the correlator output provides a measure of the visibility at two points in the (${u}$,${v}$) plane. For an East-West baseline Lz=0, and a single ellipse is centered on the (${u}$,${v}$) origin.

The characteristics and specifications
of CSRH-I are shown in Table.1. This array has 64 frequency channels and 16 channels per sub-band. The detailed specifications of CSRH-II will be introduced in further paper.
\begin{table}
\centering
\caption{CSRH-I characteristics and specifications}
\begin{tabular}{ll}
\hline\noalign{\smallskip}

      Observing band          & 0.4-2GHz \\
      Spatial Resolution      & 51.6"-10.3" \\
      Temporal Resolution     & 100 ms \\
      Dynamic range of images & $\geq (25dB)$\\
      Polarization            & left and right circular polarization \\
      frequency channel       & 64 \\
      Observing period        & 0UT to 8UT in winter, -1UT to 9UT in summer \\
      antenna efficiency      & $\geq (0.4)$ \\
      antenna noise temperature   & $\leq (120K)$ \\
\noalign{\smallskip}\hline
\end{tabular}
\label{table:first}
\end{table}
 To cover a wide frequency range with high sensitivity, the bandwidth of CSRH-I is divided into four sub bands: 0.4-0.8GHz, 0.8-1.2GHz, 1.2-1.6GHz and 1.6-2.0GHz. When the radio frequency (RF)
 signal of the Sun arrives at the analogue receiver of CSRH, it is mixed with two local oscillators. The first oscillator mixes the four sub bands at 3.6GHz, 4.0GHz, 4.4GHz and 4.8GHz
 respectively. So all the four sub bands become 2.8-3.2GHz after the operator of the first oscillator. Then, the output of the first oscillator is further mixed with the second oscillator at the
 frequency of 3.25GHz. Thus, the output frequencies of all the four sub bands range from 50MHz to 450MHz.

 In CSRH-I, to ensure all RF signals arriving simultaneously, the optic fibers transmitted the received RF signals of all channels with the same length.
 Each pair of two antennas is correlated to output a Fourier component of a solar radio image. Then we can reconstruct the brightness image through gathering all possible Fourier components from this interferometry.

\subsection{Simulation of the radiation pattern for the feed and reflector antenna system}
\begin{figure}[htb]
 \centering
   \includegraphics[width=0.55\textwidth, bb=-92 -45 588 527]{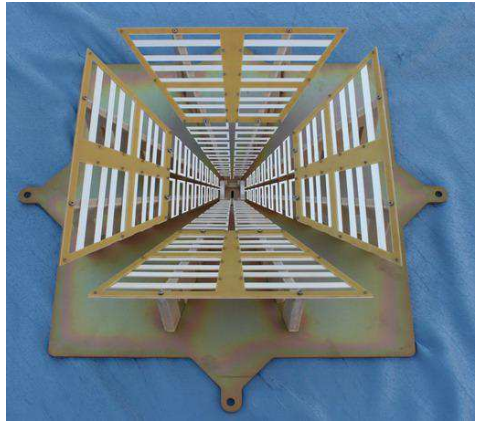}
   \caption{The photo of CSRH-I feed}
  \label{figurefeed}
\end{figure}

\begin{figure}
\centering
  \includegraphics[width=0.55\textwidth]{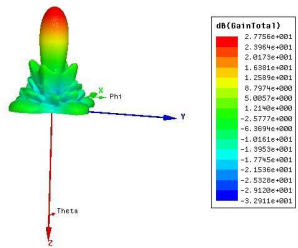}
\caption{Simulated 3D radiation pattern of feed and reflector
antenna in 0.4GHz, the right scale shows different colors
corresponding to different values of Gain total in dB}
\label{fig:2}       
\end{figure}

\begin{figure}
\centering
  \includegraphics[width=0.65\textwidth]{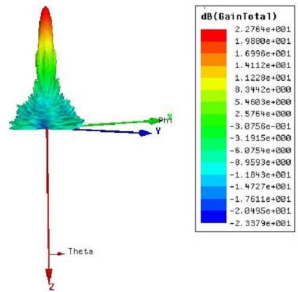}
\caption{Simulated 3D radiation pattern of feed and reflector
antenna in 1.2GHz, the right scale shows different colors
corresponding to different values of directivity total in dB}
\label{fig:3}       
\end{figure}

\begin{figure}
\centering
  \includegraphics[width=0.85\textwidth, bb=14 -10 412 228]{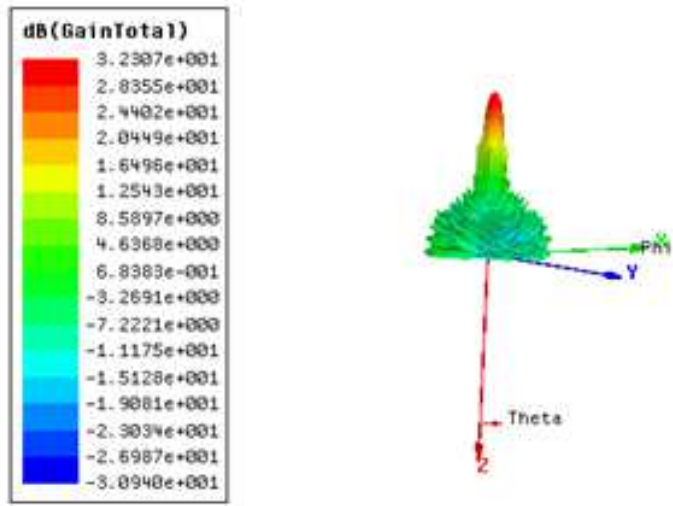}
\caption{Simulated 3D radiation pattern of feed and reflector
antenna in 2GHz, the left scale shows different colors corresponding
to different values of Gain total in dB}
\label{fig:4}       
\end{figure}

CSRH is an ultra wide band radio heliograph, in reference to choosing feed used in CSRH, there are many kinds of ultra wide band feeds (Taylor~\cite{Taylor}) in
antenna system but with some shortcomings. For example, ridged
horn is too heavy and expensive; log-periodic antenna has variable
phase center location and it spills over the edge of the dish to
reduce the gain of the antenna system. In addition, the 3dB beamwidth of the radiation patterns of these two feeds are not wide enough, so the reflector
surface cannot be fully illuminated by the feed, which will result in
the low usage of the reflector antenna. For CSRH, we use Eleven feed to receive radio signals. Because it has many advantages, such as a fixed phase
center location, well-shaped radiation patterns and low
return loss. This feed is comprised of 13 folded dipoles. The
bandwidth is decided by the ratio of the length of the "next-to-shortest" dipole to the length of the "next-to-longest" dipole. "next-to-shortest" means the shorter dipole of the two adjacent dipoles and "next-to-longest" means the longer dipole of the two adjacent dipoles.
The length of the longest dipole is given by the wavelength of the
lowest frequency. (Figure~\ref{figurefeed}) gives the photo of CSRH-I feed.
In the simulating stage, we use High Frequency Structure Simulator (HFSS) software to simulate the radiation pattern of the feed and reflector antenna system of CSRH. In order to get good radiation pattern,
dipoles in different heights vary one by one (Qing~\cite{Qing}) according
to the optimized scaling ratio. Then, the derived radiation pattern
of the feed is substituted to the feed and reflector antenna
model, we can obtain the radiation pattern of the whole system. (Figure~\ref{fig:2}), (Figure~\ref{fig:3}) and (Figure~\ref{fig:4}) show the simulated 3D radiation patterns of the feed and reflector
antenna system at 0.4GHz, 1.2GHz and 2GHz respectively. The feed is located in the focus of the parabolic reflector antenna, the pattern is symmetric about the z-axis. We can see that the radiation patterns behave a good symmetrical property on boresight.

\begin{figure}
\centering
  \includegraphics[width=0.65\textwidth]{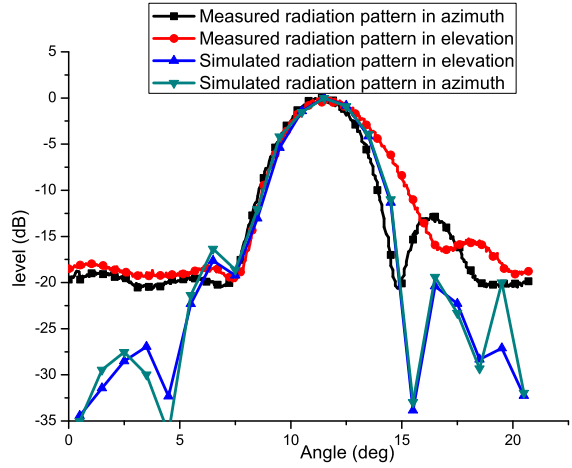}
\caption{Simulated and measured 2D radiation pattern in 1.7GHz, the black line with square and red line with circle represent radiation pattern in elevation and azimuth directions; the blue line with up triangle and the dark cyan line with down triangle represent radiation pattern in elevation direction and azimuth directions}
\label{figu:3}       
\end{figure}

 To check the consistence between the simulated and measured radiation patterns, we observed Fengyun-2E satellite at 1.7GHz. This satellite bacon signal is received by IA0 antenna. The antenna moves in up and down directions and in left and right directions, so the measured radiation patterns in elevation and azimuth directions are obtained by receiving signals coming from the maximum satellite downlink signal independently.
 The simulated and measured radiation patterns are illustrated at 1.7GHz in (Figure~ \ref{figu:3}).

 In antenna system, the parameter $ G $ is measured using two antennas(IB11 and IA0).
 IB11 was connected to a signal source, and it radiated signal to IA0, IA0 received signal from the source and the spectrum analyzer was connected with IA0. The ratio of the powers from the signal source and from the spectrum analyzer could be obtained, we could get the measured antenna gain $ G $ from (Equation~\ref{eqa:G}).
\begin{equation}\label{eqa:G}
  G = \frac{{(27000 \sim 31000)}}
{{\theta _{azimuth} \theta _{elevation} }}
\end{equation}
 where $ \theta _{azimuth}$ and $\theta _{elevation}$ are 3dB beamwidths  in azimuth and elevation directions (Kildal1~\cite{Kildal1}).
 The directivity coefficient $D$ is calculated by (Equation~\ref{eqa:Dre}).
 \begin{equation}\label{eqa:Dre}
    D = \left( {\frac{{\pi d}}
{\lambda }} \right)^2
\end{equation}
where $d$ and $\lambda$ represent the diameter of the reflector
antenna and the wavelength of the working frequency. From (Equations~
\ref{eqa:G}) and (Equations~\ref{eqa:Dre}), the measured $G$ and calculated $D$ are 34.61dB,
38.08dB respectively. Thus, the relationship between $G$, $D$ and antenna efficiency $\eta$ is expressed in Equation~\ref{E}):
   \begin{equation}\label{E}
    G = D*\eta
   \end{equation}
 then, we calculated $\eta$ is 0.45, which is satisfied with the specification of the antenna design.
\subsection{The equatorial mount and pointing error of the reflector antenna of CSRH}
To facilitate the observation of the Sun, the reflector antenna of
CSRH is designed with equatorial mount(also
called polar mount) and use the
quadrupod structure to support the
feed, the foot points of this
quadrupod are located at the edge of
reflector antenna to reduce the
influence of the plane
wave incoming. The equatorial mount is a mount for instrument that follows the rotation of the sky (celestial sphere) by having one rotational axis parallel to the Earth's axis of rotation.
In addition, it makes all the antennas
operating consistently. The monitoring subsystem sends control
signal to the operating unit of an antenna, and the antenna reacts
the signal to adjust its posture working in the best condition.

The pointing error of each antenna is
very important to an extended source
such as the quiet Sun. If the pointing
error of the antenna is not accurate,
it would influence the phase
correlation, which can not be deleted
by the post-processing. The measured
pointing error of each antenna of CSRH
is less than 9'. When all the antennas
point to the Sun after calibration, the
response of the receivers must be
uniform over the whole angular range of
the observation.

\section{Antenna noise temperature}
\label{sect:data}
\label{sec:2}
\subsection{Theoretical analysis of antenna noise temperature}
\label{sec:2.1}
In CSRH program, it is necessary to choose
better amplifier (Kildal~\cite{Kildal1}) matching the
antenna system. The antenna and receiver system noise temperature must be lower than the specification of the whole system.

The classical method of measuring antenna
noise temperature is based on detection
of astronomical source and sky
background. These measurements could be
extended for judging the system
independent of the telescope antenna.
Signals coming from the sky background
noise and astronomical source need to
be measured at frequencies supported by
the antenna-feed. With the requirement
of solar radio observation, the antenna noise
temperature(Kildal~\cite{Kildal2}) is measured
in the whole system. It relies on the
reflector configuration and the feed
radiation pattern, which means that if
the feed has low sidelobe and cross
polarization (Olsson~\cite{Olsson}), the
reflection from the ground plane will
be reduced. At the same time, it could
minimize the spillover from the
sidelobes and the selection of feed is
vital to the whole antenna system.

When observing solar radio signals, the
circular polarization (Uralov~\cite{Uralov}) is
demonstrated from studies of large
numbers of storms. The characteristic of the wave is effected by the terms of antenna
gain, effective aperture, antenna noise
temperature and so on. The reduction
of the system noise temperature could
improve the sensitivity of a radio
telescope(Li~\cite{Li}). Minimizing radio noise
temperature usually involves cooling
the amplifier from the front end of the
system. It is convenient to use cosmic
sources with small angular as a
calibrated source, because the flux
density of these sources is
already known. Regarding the Sun, it is
impossible to use a cosmic source
because the Sun is stronger than any
other sources. Thus, a noise source
as a standard source is equipped in the
input port of the analogue system. The
input port is connected to the load,
the antenna observation of sky
background and the standard noise
generator sequentially, the
output data is gathered to compute antenna noise
temperature. The thermal noise sources
used in this measurement is resistor which is
connected to receiver by coaxial lines.
The noise temperature of the receiver
and antenna system is measured by Y
factor method. The factor
Y is represented by:
\begin{equation}
    {\rm{Y  = }}10^{{\rm{(P1 - P2)/10}}}
    \label{eqa:Y}
\end{equation}
where $P1$ and $P2$ correspond to the
powers of different loads added in the
end of the system. After obtaining Y,
the antenna noise temperature $ T_A $
is computed by
\begin{equation}
   {\rm{T_A = (T_0 + Tr - Y\times Tr)/
   Y}},
   \label{eqa:Noise temperature}
\end{equation}
where $ T_0 $ represents the
ambient temperature, $ T_0 = 290K $, $ T_r $ means the temperature of the analogue receiver.

The noise temperature of the whole
system is the antenna system noise
temperature and receiver noise
temperature. (Equation~
\ref{eqa:P50}),(Equation~\ref{eqa:Pns}) and (Equation~\ref{eqa:Noise
Y}) show different received powers with
different loads. In these
equations,
$k=1.3806505\times10^{(-23)}$ J/K, B
represents the bandwidth, $ G_r $ is the
gain of the system coming from the
noise source of the receiving
system. the noise temperature $ T_{NS}
$ means terminal connecting with the
noise source, $ T_{50} = 290K $, the noise temperature $ T_{50} $ means terminal
connecting with standard impedance
50$\Omega$,.
\begin{equation}
  \left[ {\left( {{\rm{T}}_{{\rm{50}}}  + {\rm{T}}_{\rm{r}} } \right){\rm{kB}}} \right] + \left[ {{\rm{G}}_{\rm{r}} } \right] = \left[ {{\rm{P}}_{{\rm{50}}} } \right]
  \label{eqa:P50}
\end{equation}
\begin{equation}
  \left[ {\left( {{\rm{T}}_{{\rm{NS}}}  + {\rm{T}}_{\rm{r}} } \right){\rm{kB}}} \right] + \left[ {{\rm{G}}_{\rm{r}} } \right] = \left[ {{\rm{P}}_{{\rm{NS}}} } \right]
  \label{eqa:Pns}
\end{equation}
\begin{equation}
\left[ {\left( {{\rm{T}}_{\rm{A}}  + T_r } \right){\rm{kB}}} \right] + \left[ {{\rm{G}}_{\rm{r}} } \right] =  \left[ {{\rm{P}}_{{\rm{SKY}}} } \right]
\label{eqa:Noise Y}
\end{equation}
The square brackets in (Equation~\ref{eqa:P50}), (Equation~\ref{eqa:Pns}), (Equation~\ref{eqa:Noise Y}) mean that the unit of each value is dB. From (Equation~\ref{eqa:Tr}), (Equation~\ref{eqa:Gr}), (Equation~\ref{eqa:TA}), we could get the representations of $ T_r $, $ G_r $ and $ T_A $ respectively.
\begin{equation}\label{eqa:Tr}
  T_r  = \frac{{T_{NS}  - T_{50} \left( {\frac{{P_{NS} }}{{P_{50} }}} \right)}}{{\frac{{P_{NS} }}{{P_{50} }} - 1}}
\end{equation}
\begin{equation}\label{eqa:Gr}
   G_r  = \frac{{P_{50} \left( {\frac{{P_{NS} }}{{P_{50} }} - 1} \right)}}{{kB\left( {T_{NS}  - T_{50} } \right)}}
\end{equation}
\begin{equation}\label{eqa:TA}
   T_A  = \frac{{P_{sky} }}{{P_{50} }}(T_{50}  + T_r ) - T_r
\end{equation}

\subsection{The measured results of CSRH}
For CSRH radio heliograph, we test the
gain by using Y factor method (Penzias~\cite{Penzias}) in the
whole system. (Table~\ref{tab:noise
source}) gives the measured noise
temperatures of the noise source at
different frequencies. The relationship between Excess Noise Ratio(ENR) and the noise source temperature is:
\begin{equation}\label{1}
{\rm{ENR = }}\frac{{{T_{NS}} - {T_0}}}{{{T_0}}}
\end{equation}
Where, ${T_0}$ is the ambient temperature, $ {{T_0} = 290K }$. From (Table~
\ref{tab:noise source}), it can be
observed that the noise temperature
decreases as the frequency increases.
(Table~\ref{tab:noise power}) lists the
noise powers of different
terminals: noise source, 50$\Omega$
impedance, sky background and the Sun.
Based on the measured results of
(Table~\ref{tab:noise power}), we can calculate
system noise ($Tr$), system Gain ($ Gr
$) and antenna noise temperature ($ T_A
$) by using (Equation~\ref{eqa:Tr}),
(Equation~\ref{eqa:Gr}), and (Equation~\ref{eqa:TA})
respectively. The calculated results
are shown in (Table~\ref{tab:Noise}). The fourth column of this table gives the measured antenna noise temperature, they all less than the specification 120K in different frequencies.
\begin{table}
\caption{The noise temperatures
($T_{NS}$) of the noise source at
different frequencies, ENR means the multiple number of the noise source above the ambient temperature, this value could be tested according to an instrument.}
 \centering
\begin{tabular}{clc}

 \hline\noalign{\smallskip}
Radio frequency(GHz) & ENR(dB) & Noise temperature(K) \\
\noalign{\smallskip}\hline\noalign{\smallskip}
  0.75  & 21.28 & 392230.18 \\
  0.8   & 21.03 & 37051.90    \\
  1.025 & 20.57 & 33357.24 \\
  1.2   & 20.18 & 30517.21    \\
  1.45  & 19.82 & 28112.62 \\
  1.95  & 18.71 & 21837.56   \\
\noalign{\smallskip}\hline
\label{tab:noise source}
\end{tabular}
\end{table}
\begin{table}
\centering
\caption{The noise powers of the
different terminals including noise
generator, 50$\Omega$ impedance, sky
background and the Sun at different
frequencies ($ P_{NS} $, $P_{50}$, $
Psky $, and $Psun$ represents the
measured powers of the terminals.)}
\begin{tabular}{clclc}
\hline\noalign{\smallskip}
Radio frequency(GHz) & $ P_{NS} $(dBm) & $P_{50}$(dBm) &$ Psky $ (dBm) & $Psun$ (dBm)\\
\noalign{\smallskip}\hline\noalign{\smallskip}
  0.75  & -26.19 & -44.93 & -46.58 & -39.25 \\
  0.8   & -24.98 & -42.81 & -44.42 & -37.83 \\
  1.025 & -22.21 & -40.52 & -42.78 & -35.65 \\
  1.2   & -28.31 & -44.6  & -46.36 & -41.5  \\
  1.45  & -29.81 & -46.80 & -48.57 & -42.13 \\
  1.95  & -28.86 & -43.61 & -45.2  & -41.5  \\
\noalign{\smallskip}\hline
\end{tabular}
\label{tab:noise power}

\end{table}
\begin{table}
\centering
\caption{The measured system noise
($Tr$), system Gain ($Gr$) and antenna
noise temperature ($T_{A}$) at
different frequencies}
\begin{tabular}{clcl}
\hline\noalign{\smallskip}
Radio frequency(GHz) & $T_r (K)$ & $G_r (dB)$ & $T_A(K)$  \\
\noalign{\smallskip}\hline\noalign{\smallskip}
 0.75  & 268.3 & 61.7 & 92.9 \\
  0.8   & 326.05& 63.12& 99   \\
  1.025 & 235.7 & 66.4 & 58.9 \\
  1.2   & 437.3 & 60.61 & 48\\
  1.45  & 308.3 & 59.9 & 69.8 \\
  1.95  & 456.42 & 61.49 & 61.17 \\
\noalign{\smallskip}\hline
\label{tab:Noise}
\end{tabular}

\end{table}
\section{Data processing for CSRH-I}
\label{sec:3}
\subsection{Solar radio burst observation}
\begin{figure*}[htb]
\centering
\subfloat[Observed fringe included the sin and cosine components]{\includegraphics[width=4.25in]{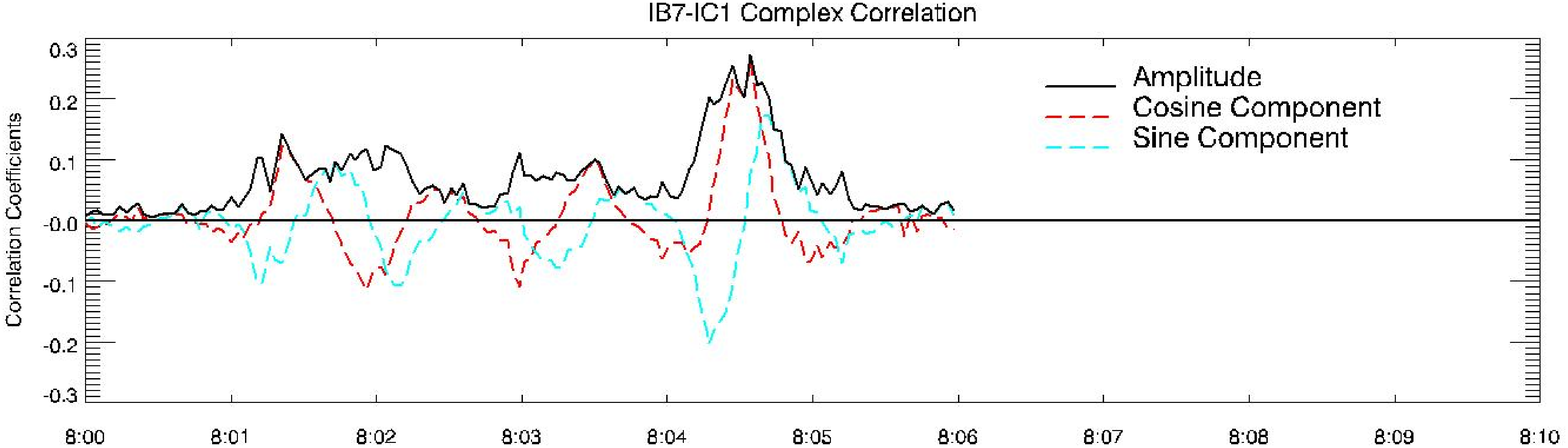}%
\label{a}}
\hfil
\subfloat[The flux density observed by GOES]{\includegraphics[width=4.35in]{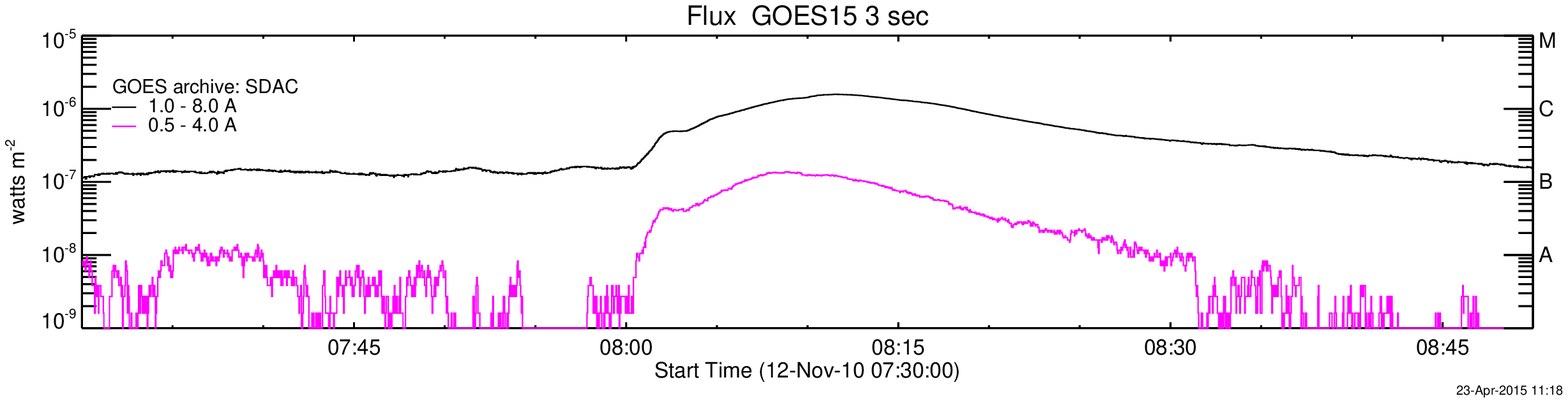}%
\label{b}}
\hfil
\subfloat[The measured result obtained by IC1 ]{\includegraphics[width=4.55in]{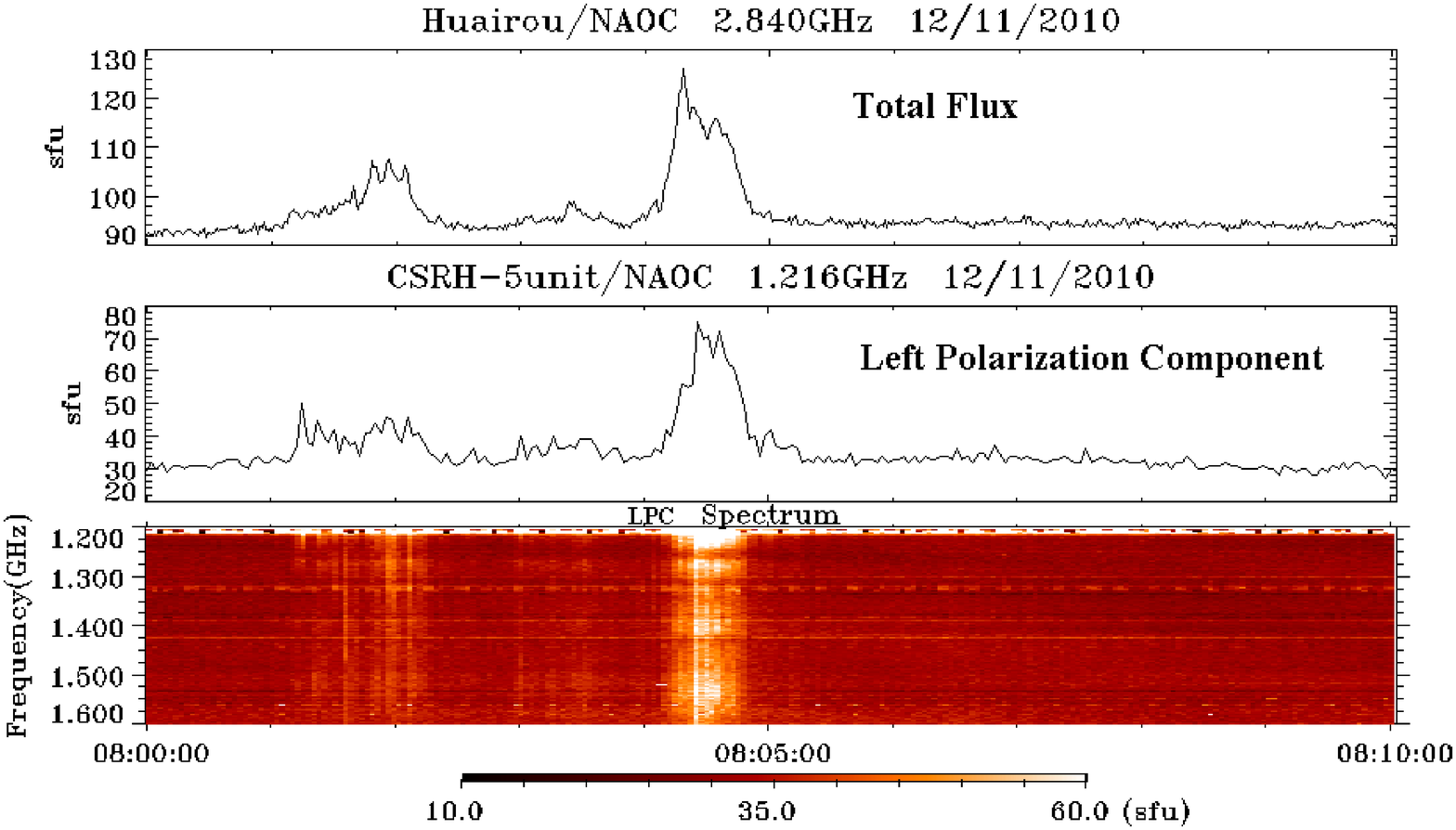}%
\label{c}}
\caption{The comparison between different instruments observations}
\label{solar}
\end{figure*}

 A solar radio burst instance is
provided to validate the effectiveness
of CSRH, this event occurred on November 12,
2010. By using a 5-element system of
CSRH-I, we successfully observed a solar radio burst with an associated X-ray flare C1.5
class. In
Huairou observatory station,  SBRS also observed
this instance at the same time. Another instrument called GOES also observed this event in X-ray at that time, (Figure~
\ref{solar}) (a) gives the observed
fringe with IB7-IC1 baseline, IB7 and IC1 are two antennas of CSRH-I, the black curve shows the amplitude of the
fourier component, the red and blue curves show cosine and sin components;
(Figure~\ref{solar}) (b) gives the
flux density of GOES at wavelength of 1-8 Angstrom showing in black line
and 0.5-4 Angstrom drawing in red line; (Figure~\ref{solar}) (c) shows the total flux density of Huairou instrument observation and IC1 observed result.

\subsection{Satellite image using Aperture synthesis method}
\begin{figure*}
\centering
  \includegraphics[width=0.75\textwidth]{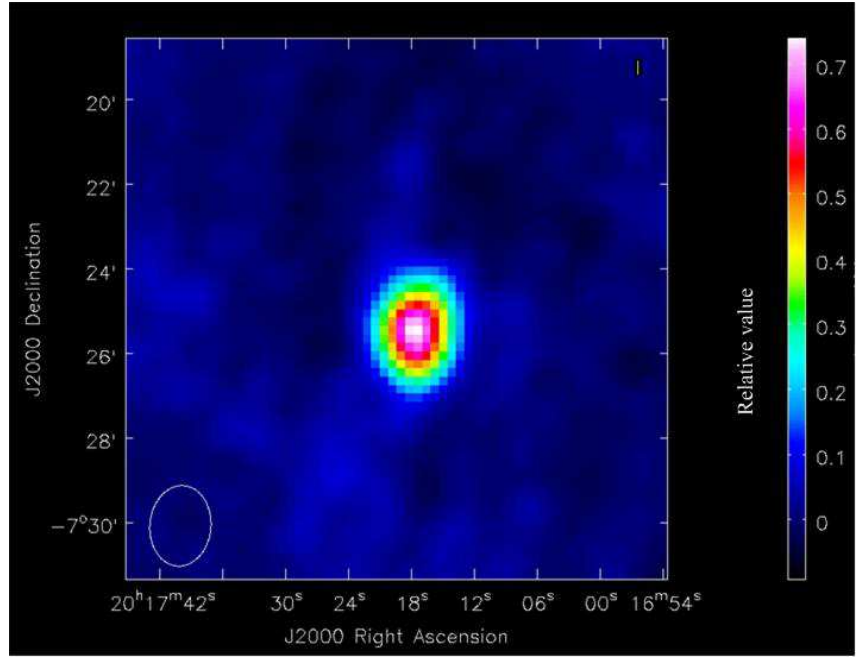}
\caption{Stokes parameter I: the
cleaned observed image of Fengyun-2E
satellite at 1.7GHz on June 4th, 2013, the
ellipse in the lower left corner is the
synthesized beam, the vertical scale
bar represents the relative level of
the background}
\label{fig:sate2}       
\end{figure*}

 In reference to CSRH-I image calibration, the diameter of the antenna is too small to use non-solar compact sources as calibrating sources. Therefore, we use Fengyun-2E satellite source as a point source in the beginning stage of CSRH establishment at 1.7GHz. The observed result of Fengyun-2E is
 drawn in (Figure~\ref{fig:sate2}), there are totally 930 fourier components using 31 antennas, each pair of these antennas is interfered to provide a fourier component of the observed source. The brightness image of the observed source can be obtained (Wang~\cite{Wang}) through applying inverse fourier transform to the gathered fourier components.

For detecting satellite source,
observation is obtained during the pass
of the source through a stationary
beam. (Figure~\ref{fig:sate2}) shows the Stokes
parameter I, which represents the total
intensity of the satellite source. The
phase error of spherical waves coming
from satellite source would be more
than $ {360^\circ } $ for the long
baseline, and varies with the motion of
the satellite. However, such error
could be removed successfully using two observations. (Figure~\ref{fig:sate2})
is obtained from CASA, the integration time is 30ms
and the beam size of this image corresponding to the antenna
beam is 1.25'*2.489'. This observation
shows the good performance of the
imaging capability of CSRH-I.
\section{Conclusion}
\label{sec:4}
 The characteristics of CSRH are presented in this paper. From these data, we could know that the system gain $Gr$ is larger than 60dBi and antenna noise temperature $T_A$ is less than $120K$
 respectively, which are satisfied with the specifications of our science requirement. After some of the antennas installed in Mingantu Observatory, a solar radio burst at 1.302GHz was successfully captured by CSRH 5-element on 2010, November 12, at the same time, SBRS and GOES also observed this event. Another result is about the point source of satellite image at 1.7GHz ,which shows the imaging ability of this system. CSRH will establish solar radio images in decimeter and centimeter wavelength more accurately.
\begin{acknowledgements}
We acknowledge the anonymous referee for his/her valuable suggestions and comments of this paper. The author also wish to thank Long Xu, Chong Huang, Cheng Ming Tan, Jing Huang, for their lots of advice on this paper. This work is supported by NSFC grants (Nos.: MOST2011CB811401£¬11221063, 10778605, 11003028, 11203042, U1231205), National Major Scientific Equipment Research and Design project (ZDYZ2009-3).
\end{acknowledgements}


\begin{thebibliography}{99}

 \bibitem[1996]{Takano} Takano T., Springer, Netherlands, 1996, 569

 \bibitem[1998]{Bastian} Bastian T.S., Benz, A.O., Gary, D.E., 1998, \araa, 36, 131

 \bibitem[2014]{Chen} Chen J.J., Bai Z.R., Luo A.L., Zhao Y.H., 2014, RAA, 15, 607

 \bibitem[1965]{Penzias} Penzias A. A., Wilson, R. W., 1965, The Astronomical Journal, 142, 1149

 \bibitem[2005]{Aghdam} Aghdam K. M. P., Farajj-Dana R., Rashed-Mohassel J., 2005, IEEE Proc. Microw. Antennas Propag., 59,392

 \bibitem[2007]{Gary} Liu Z., Gary D. E., Nita G. M., White S. M., Hurford G. J., 2007, Publications of the Astronomical Society of the Pacific, 119, 307

 \bibitem[1998]{Uralov} Uralov A. M., Grechnev V. V., Lesovoi S. V., Sych R. A., Kardapolova N. N., Smolkov G. Ya, Treskov T. A., 1998, Solar Physics, 178, 119

 \bibitem[2009]{Kildal1} Kildal P. S., 2009, Chalmers University, Chalmers, 44

 \bibitem[1995]{Kildal2} Kildal P. S., Sipus, Z., 1995, IEEE Proc. Microw. Antennas Propag., 37,114

 \bibitem[1994]{Nakajima} Nakajima H., Nishio M., Enome S., 1994, IEEE Proc. Microw. Antennas Propag.,, 82,705

 \bibitem[2006]{Olsson} Olsson R., Kildal P. S., Weinreb S., 2006, IEEE Proc. Microw. Antennas Propag.,, 54,368

 \bibitem[1999]{Qing} Qing X. M., Chia Y. W. M. 1999, Electron. Lett., 35,2154

 \bibitem[2008]{Yan} Yan Y., Zhang J., Wang W., Liu F., Chen Z., Ji G., 2008, Earth, Moon, and Planets, 104,1

 \bibitem[2014]{Les} Sergey V. Lesovoi, Alexander T. Altyntsev, Eugene F. Ivanov, Alexey V. Gubin, 2014, RAA, 864

 \bibitem[2013]{Wang} Wang W., Yan Y. H., Liu D. H., Chen Z. J., Su C., Liu F., Geng L. H., Chen L. J., Du J., 2013, PASJ, 65, sp18-1

 \bibitem[2015]{Li} Li S., Yan Y. H., Chen Z. J., Wang W., Zhang F.S., 2015, PASA, 32, 1

 \bibitem[1999]{Taylor} Taylor G. B., Carilli C. L., Perley R. A., 1999, ASP Conf. Ser. v. 11

 \bibitem[2008]{Lindsay} Lindsay D. Mi, Howell C, K., 2008, The Journal of the Astronautical Sciences, 56,71

%
\end{thebibliography}
\end{document}